\documentstyle[12pt,epsfig]{article}
\begin{document}

\begin{flushright}
{\bf UCD--95--9}
\end{flushright}

\vspace{.5cm}

\begin{center}

{\Large\bf
QCD Corrections to Diboson Production}\footnote{Contribution
to the Proceedings of the International Symposium on Vector Boson
Self-Interactions, UCLA, Feb. 1--3, 1995.}

\vspace{.7cm}

{\bf J.~Ohnemus}\\[1mm]
{\it Department of Physics}\\
{\it University of California}\\
{\it Davis, CA 95616}

\end{center}

\vspace{.7cm}

\begin{abstract}
The QCD radiative corrections to hadronic diboson production are reviewed.
The radiative corrections for $W^{\pm}\gamma$, $Z\gamma$,
$ZZ$, $W^+W^-$, and $W^{\pm} Z$ are discussed.
Similarities and differences in the behavior of the order
$\alpha_s$ cross sections for these processes are emphasized.
\end{abstract}

\section*{Introduction}

The production of weak boson pairs is an important topic to study at
hadron colliders because these processes can be used to test the
standard model (SM) as well as probe beyond it \cite{FIRSTWZ}.
Diboson production is important for the following reasons.

\begin{itemize}

\item{} The $W^{\pm}\gamma$, $W^{\pm}Z$, and $W^+ W^-$ processes can
be used to test the trilinear $WW\gamma$ and $WWZ$ couplings. These
couplings are completely fixed by the ${\rm SU(2)} \otimes {\rm U(1)}$
gauge structure of the SM, thus measurements of these couplings
provide stringent tests of the SM.   Remarkable progress has recently
been made in measuring these couplings at the Fermilab Tevatron
collider \cite{UCLA}.

\item{} The electroweak symmetry breaking (EWSB) mechanism can be
probed  by studying weak boson pair production.  The EWSB mechanism is
unknown, but it is believed that either there exists a scalar particle
with mass $m < 1$~TeV or else the longitudinal components of the
$W$ and $Z$ bosons become strongly interacting for parton
center-of-mass energies larger that about 1~TeV \cite{EWSB}.  For
example, the observation of resonance production of $ZZ$, $W^+ W^-$,
or $\gamma \gamma$ would be a signal for the standard model Higgs
boson, whereas enhanced production of longitudinally polarized $W$ and
$Z$ pairs would be evidence for a strongly interacting EWSB scenario.

\item{} Diboson production is a potential background to new physics.
New heavy particles, such as $H^0$, $H^{\pm}$, $\rho_{\rm TC}^{}$,
$\eta_{\rm TC}^{}$, $W^\prime$, $Z^\prime$, $\tilde q$,  and $\tilde
g$ can decay into weak boson pairs.

\end{itemize}

In order to test and probe the SM with hadronic diboson production,
it is necessary to have precise calculations of SM diboson production,
which means the cross sections must be calculated to
next-to-leading-order (NLO). The NLO cross section is, in general,
less sensitive to the choices of the arbitrary factorization and
renormalization scales.

The results described here are based on complete ${\cal O}(\alpha_s)$
calculations of the  processes $p\,p\hskip-7pt\hbox{$^{^{(\!-\!)}}$}
\rightarrow  V_1 V_2 + X$ where $V_i = W, Z, \gamma$ \cite{NLOVV}. The
calculations also include the leptonic decays  of the $W$ and $Z$
bosons \cite{NEWJO,BHO}. This is an important feature to include since
the $W$ and $Z$ bosons are observed experimentally via their leptonic
decay products. It is therefore important  to include the experimental
cuts on the decay leptons when comparing a theoretical calculation to
the experimental data.

The calculations have been done using a combination of analytic and
Monte Carlo integration techniques.  Among the advantages of this
formalism are:

\begin{itemize}
\item{} It is easy to impose cuts in the calculation.

\item{} It is possible to calculate any number of observables
simultaneously by simply histogramming the quantity of interest.

\item{} It is possible to calculate not only the NLO inclusive cross
section, but also the 0-jet and 1-jet exclusive cross sections.
\end{itemize}

Details of the formalism can be found in the
original references \cite{NLOVV,NEWJO,BHO}.

\section*{The $Z\gamma$ and $W\gamma$ Processes}
The first processes to be considered are the $Z\gamma$ and $W\gamma$
processes. The total LO and NLO cross sections for these processes are
plotted as functions of the center of mass energy in Fig.~1.   The
difference between the NLO and LO curves is the ${\cal O}(\alpha_s)$
correction. In the $Z\gamma$ process, the ${\cal O}(\alpha_s)$
corrections range from 10\% to 30\% over the domain of $\sqrt{s}$.
This is what one naively expects since $\alpha_s$ is of order 0.10.
In the $W\gamma$ process, on the other hand, the corrections range
from 20\% at small $\sqrt{s}$ to a surprising  300\% at large
$\sqrt{s}$.

In order to understand the large ${\cal O}(\alpha_s)$  corrections in
the $W\gamma$ process, it is instructive to compare the behavior of
the $2 \to 2$ and $2 \to 3$ processes for $Z\gamma$ and $W\gamma$
production.   Figure~2(a) compares the $2\to 2$ cross sections.
Normally, hadronic $W$ production is about twice as large as hadronic
$Z$ production because the  $W$-to-quark coupling is about twice as
big as the $Z$-to-quark coupling.   However, for the $W\gamma$ and
$Z\gamma$ processes, exactly the opposite behavior is seen; the
$W\gamma$ cross section is only half as big as the $Z\gamma$ cross
section. The $W\gamma$ cross section is smaller because it is
suppressed by a radiation amplitude zero (RAZ) \cite{RAZ}.  Delicate
cancellations in the $W^{\pm}\gamma$ amplitude cause it to vanish at
$\cos\theta^* = \pm {1\over 3}$ where $\theta^*$ is the parton
center-of-mass scattering angle.

The $2 \to 3$ cross sections for $W\gamma$ and $Z\gamma$ are compared
in Fig.~2(b). Here a jet is defined as a final state quark or gluon
with transverse momentum $p_T^{} > 50$~GeV and pseudorapidity $|\eta|
< 3$. The cross sections have been decomposed into contributions from
$qg$  and $q\bar q$ initial states ($qg$ also includes $\bar q g$).
The $q g \to W\gamma + 1$~jet cross section is about twice as big as
the $qg \to Z\gamma + 1$~jet cross section, as naively expected.  (The
$q g \to W\gamma q$ subprocess does not have a RAZ.) The $q \bar q \to
W\gamma + 1$~jet and  $q \bar q \to Z\gamma + 1$~jet cross sections,
on the other hand, are nearly equal, indicating that the  former is
still suppressed relative to the later. (The $q \bar q \to W\gamma g$
subprocess has a RAZ in the limit $E_g \to 0$.)

In summary, the $2 \to 2$ $W\gamma$ cross section is suppressed
relative to the $2 \to 2$ $Z\gamma$ cross section by a RAZ,  while the
$2 \to 3$ $W\gamma$ cross section is larger than the $2\to 3$
$Z\gamma$ cross section  due to the larger $W$-to-quark coupling.  The
net result of these two behaviors  is that the ${\cal O} (\alpha_s)$
corrections  are much larger for $W\gamma$ production than for
$Z\gamma$ production.

Figure~3 again shows the total $Z\gamma$ and $W\gamma$  cross sections
versus $\sqrt{s}$, but now the NLO cross sections have been decomposed
into the Born cross sections and  ${\cal O} (\alpha_s)$ corrections
from $q \bar q$ and $qg$ initial states.  This decomposition shows
that the  ${\cal O} (\alpha_s)$ $q\bar q$ corrections tend to be
proportional to the Born cross section, whereas the  ${\cal O}
(\alpha_s)$ $qg$ corrections increase rapidly with  $\sqrt{s}$.   The
${\cal O} (\alpha_s)$ $qg$ corrections increase with  $\sqrt{s}$
because the gluon density increases with $\sqrt{s}$.

Figure~4 shows the $p_T^{}(\gamma)$ spectra for $Z\gamma$ and $W^+
\gamma$ production at the Large Hadron Collider (LHC) center of mass
energy  ($\sqrt{s} = 14$~TeV).  The figure shows that the NLO
corrections increase  with $p_T^{}(\gamma)$.  This behavior is common
to all the diboson processes; the NLO corrections increase with the
$p_T^{}$ of the boson.

The rapidity distribution of the photon in the diboson rest frame is
shown in Fig.~5 for the Tevatron center of  mass energy ($\sqrt{s} =
1.8$~TeV).   For the $Z\gamma$ process, the distribution exhibits the
usual bell-shaped rapidity distribution, however, for the $W\gamma$
process, the distribution has a pronounced dip in the central rapidity
region.  This dip is due to the RAZ in the $W\gamma$ process.   At the
Tevatron energy, the NLO corrections slightly fill the dip, but do not
obscure it. Figure~6 shows the photon rapidity distribution at the LHC
energy.  The NLO corrections are now very large in the $W\gamma$
process and they completely fill the dip in the central rapidity
region. It may still be possible, however, to observe the dip in the
$W\gamma + 1$~jet exclusive cross section \cite{BHO}.

Figure~7 compares the $p_T^{}(\gamma)$ spectra for the $Z\gamma$ and
$W\gamma$ processes at the Tevatron energy. This comparison shows that
at high $p_T^{}(\gamma)$,  the $W\gamma$ distribution falls more
rapidly than the $Z\gamma$ distribution.  This behavior is also due to
the  RAZ in the $W\gamma$ process.

\section*{The $ZZ$, $W^+W^-$, and $WZ$ Processes}

Attention now turns to the $ZZ$, $W^+ W^-$, and $WZ$ processes. The
transverse momentum distributions for these  processes are shown in
Fig.~8. The figure shows that  the NLO corrections increase with the
$p_T^{}$ of the weak boson and are quite large at high values of
$p_T^{}$. Also note that the NLO corrections increase in the order
$ZZ$, $W^+W^-$, $WZ$.  This behavior will be discussed later.

Figure~9 again shows the $p_T^{}$ spectra of the weak bosons, but now
the 0-jet and 1-jet exclusive components of the NLO inclusive cross
section are also shown. (The 0-jet and 1-jet exclusive cross sections
sum to the NLO inclusive cross section.) This decomposition shows that
the bulk of the large corrections at high $p_T^{}$ are due to events
containing a hard jet in the final state. The jet definition used here
is $p_T^{}(jet) > 50$~GeV and  $|\eta(jet)| < 3$.

The large enhancements to the cross section at high $p_T^{}$ can be
traced to collinear splittings in diagrams such as $q g \to Z q$
followed by $q \to q W$; the $Z$ and the quark are produced with high
$p_T^{}$ and  the quark subsequently radiates a nearly collinear $W$.
In the collinear limit, the $q g \to WZq$ subprocess can be
approximated by \cite{FRIXWZ}
\begin{equation}
d\sigma(qg \rightarrow WZq) \approx d\sigma(qg \rightarrow Zq) \,
{g^2 \over 16 \pi^2} \, \log^2 \left( {p_T^2(Z) \over M_W^2} \right) \>.
\end{equation}
Figure~10 compares this collinear approximation to the full NLO
calculation and shows that the approximation describes well the shape
of the $p_T^{}$ distribution at high $p_T^{}$.

The scale dependance of the total $WZ$ cross  section is illustrated
in Fig.~11. A common scale $Q$ has been used for both the
renormalization scale $\mu$ and the factorization scale $M$.  The Born
and NLO inclusive cross sections are shown along with the 0-jet and
1-jet components of the NLO inclusive  cross section. The 1-jet cross
section is a LO quantity and thus has considerable  scale dependance.
The 0-jet cross section, on the other hand, is a NLO quantity and
exhibits  little scale dependance.  The decomposition shows that  the
scale dependance of the NLO inclusive cross section is dominated by
the scale dependance of the 1-jet component.

Figure~12 compares the $p_T^{}$ spectra of the weak bosons for the
$ZZ$, $W^+W^-$, and $WZ$ processes. The $ZZ$ and $W^+W^-$
distributions have the same shape at high $p_T^{}$ and are parallel to
one another, whereas the $WZ$ distribution falls more rapidly. A
similar behavior was observed earlier in Fig.~7 where the $Z\gamma$
and $W\gamma$ processes were compared. In the present case, the $WZ$
$p_T^{}$ spectrum falls faster than the $ZZ$ and $W^+ W^-$ spectra
because of an approximate amplitude zero \cite{UJH} in the $WZ$
process.

\subsection*{Approximate Amplitude Zero}

The $q_1 \bar q_2 \to WZ$ subprocess is very similar to the  $q_1 \bar
q_2 \to W\gamma$ subprocess, in fact, they are described by the same
set of Feynman diagrams, with $Z$ and $\gamma$ interchanged. Recall
that the RAZ in the $W\gamma$ process gave rise to a large  ${\cal
O}(\alpha_s)$ correction. A difference between the two processes is
that  whereas the $W^{\pm}\gamma$ process has an exact  amplitude zero
at $\cos\theta^* = \pm {1\over 3}$, the $W^{\pm}Z$ process has only an
approximate amplitude zero at $\cos\theta^* = \pm 0.1$.  Basically,
what happens in the $WZ$ case is that the dominant helicity amplitudes
have an exact zero, while the other helicity amplitudes remain finite
but small. The approximate amplitude zero in the $WZ$ process causes
the  NLO corrections to be larger than they were in either the $ZZ$ or
$W^+ W^-$ processes.  The approximate amplitude zero suppresses the
$WZ$ Born cross section and thus makes the NLO corrections appear
large. A more in depth discussion of approximate amplitude zeros can
be  found in the talk by T.~Han \cite{HAN}.

\section*{Summary}

The QCD radiative corrections to  weak boson pair production  at
hadron colliders has been reviewed.  The ${\cal O} (\alpha_s$) cross
sections for the diboson combinations $Z\gamma$, $W\gamma$, $ZZ$, $W^+
W^-$, and $WZ$ have been discussed and compared. Some general features
of the  ${\cal O}(\alpha_s)$ cross sections are summarized here.

\begin{itemize}

\item{} The NLO corrections increase with the center-of-mass energy.
This is due to the opening of the $q g \to V_1 V_2 q$ subprocess at
${\cal O} (\alpha_s)$ in conjunction with the  gluon density which
increases with the center-of-mass energy.

\item{} The NLO corrections are largest at high $p_T^{}(V)$. This is
due to collinear splittings in the $q g \to V_1 V_2 q$ subprocesses
which give rise to an enhancement factor $\log^2(p_T^2(V_1)/M_2^2)$.

\item{} The bulk of the large corrections at high $p_T^{}(V)$ come
from events which contain a hard jet in the final state.

\item{} $p_T^{}$ distributions are most affected by the NLO
corrections. These distributions tend to be enhanced at large values
of $p_T^{}$.

\item{} Invariant mass and angular distributions under go relatively
little change in shape at NLO, instead, these distributions tend to be
scaled up uniformly.

\item{} The NLO corrections to $W\gamma$ production are large due to a
radiation amplitude zero.

\item{} The NLO corrections to $WZ$ production are large due to an
approximate amplitude zero.

\item{} The NLO corrections are modest at the Tevatron center of mass
energy but are significant at the LHC energy.

\end{itemize}

\begin{figure} 
\psfig{file=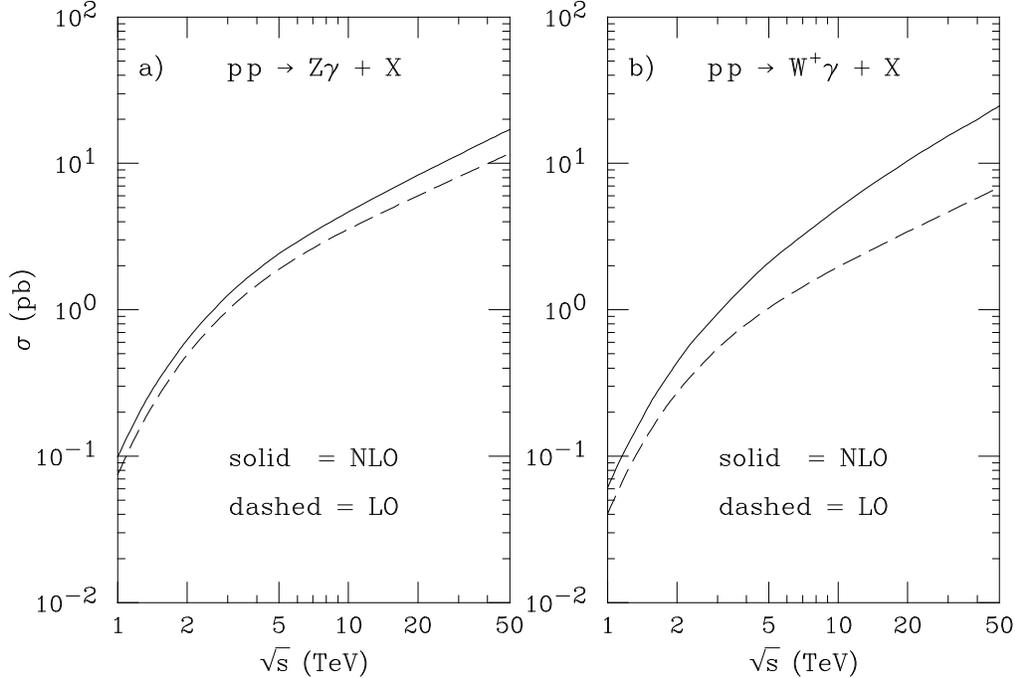,width=6.0in,clip= }
\caption{Total cross section as a function of the center-of-mass energy for
(a) $p p \to Z\gamma + X$ and (b) $pp \to W^+ \gamma + X$.
The LO and NLO cross sections are shown.}
\end{figure}

\begin{figure} 
\psfig{file=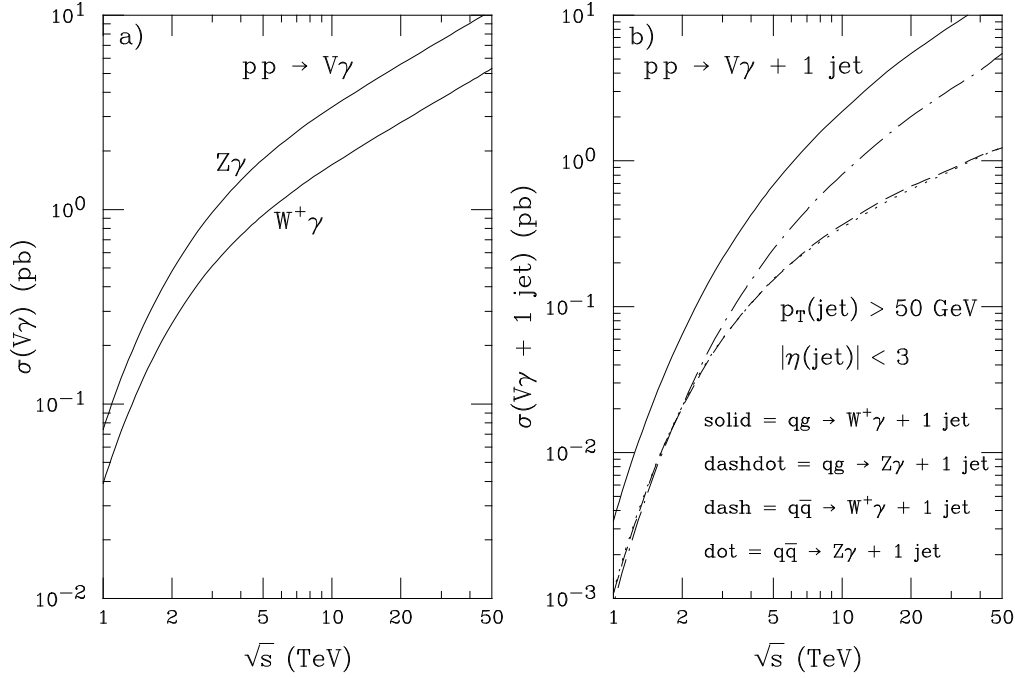,width=6.0in,clip= }
\caption{(a) The $2 \to 2$ Born cross sections for $pp \to Z\gamma$
and $p p \to W^+ \gamma$.
(b) The $2 \to 3$ cross sections for $Z\gamma$ and $W^+ \gamma$ production.
The cross sections have been decomposed into contributions from $q \bar q$
and $qg$ initial states.}
\end{figure}

\begin{figure} 
\psfig{file=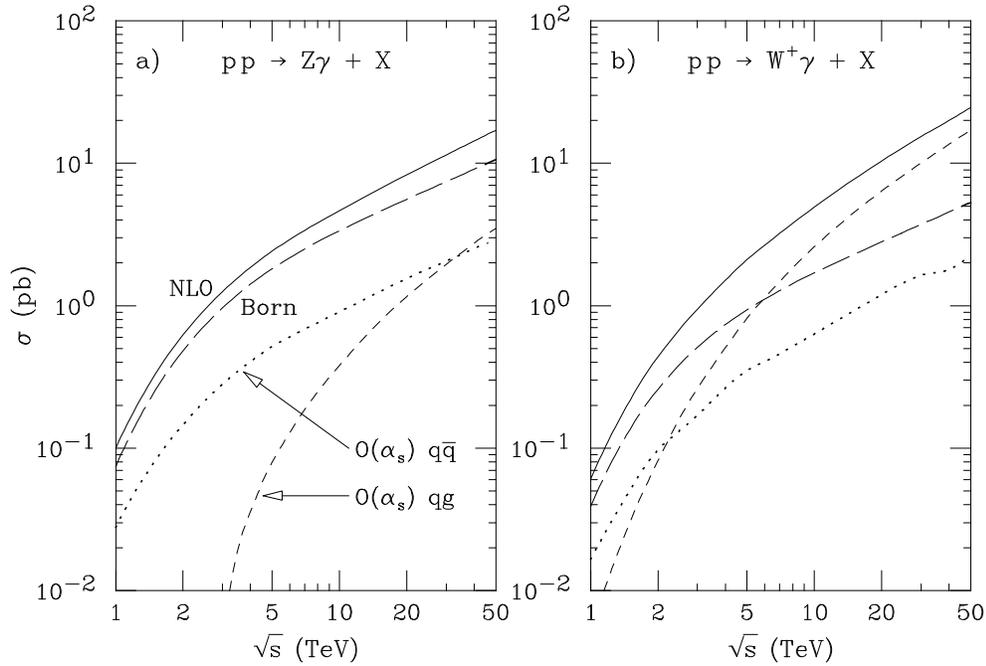,width=6.0in,clip= }
\caption{Same as Fig.~1, but now the NLO cross section has been
decomposed into the Born cross section
and the order $\alpha_s$
corrections from $q\bar q$ and $qg$ initial states.}
\end{figure}

\begin{figure} 
\psfig{file=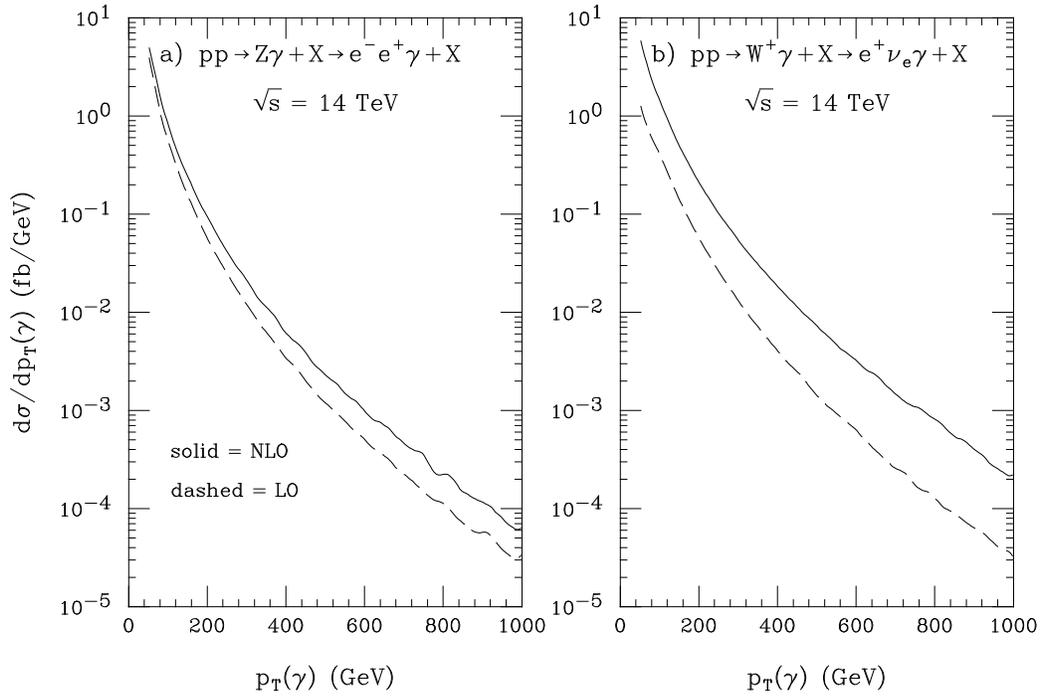,width=6.0in,clip= }
\caption{Photon transverse momentum distributions at the LHC energy for
(a) $p p \to Z   \gamma + X \to e^- e^+   \gamma + X$ and
(b) $p p \to W^+ \gamma + X \to e^+ \nu_e \gamma + X$}.
\end{figure}

\begin{figure} 
\psfig{file=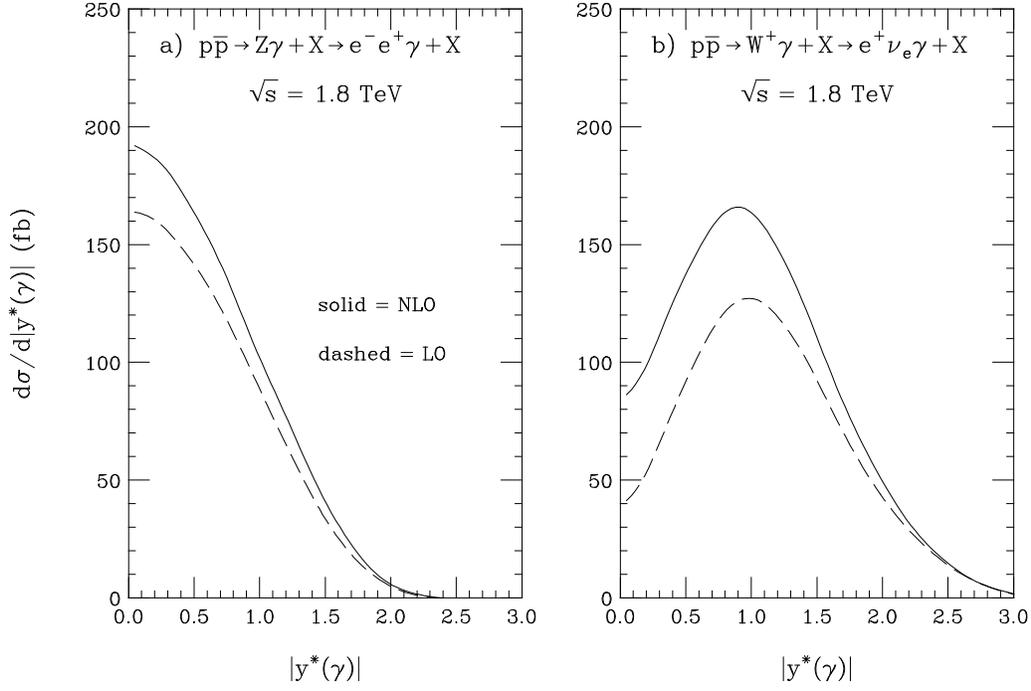,width=6.0in,clip= }
\caption{Photon rapidity distributions in the diboson rest frame
at the Tevatron energy for
(a) $Z\gamma$ production and
(b) $W^+ \gamma$ production.}
\end{figure}

\begin{figure} 
\psfig{file=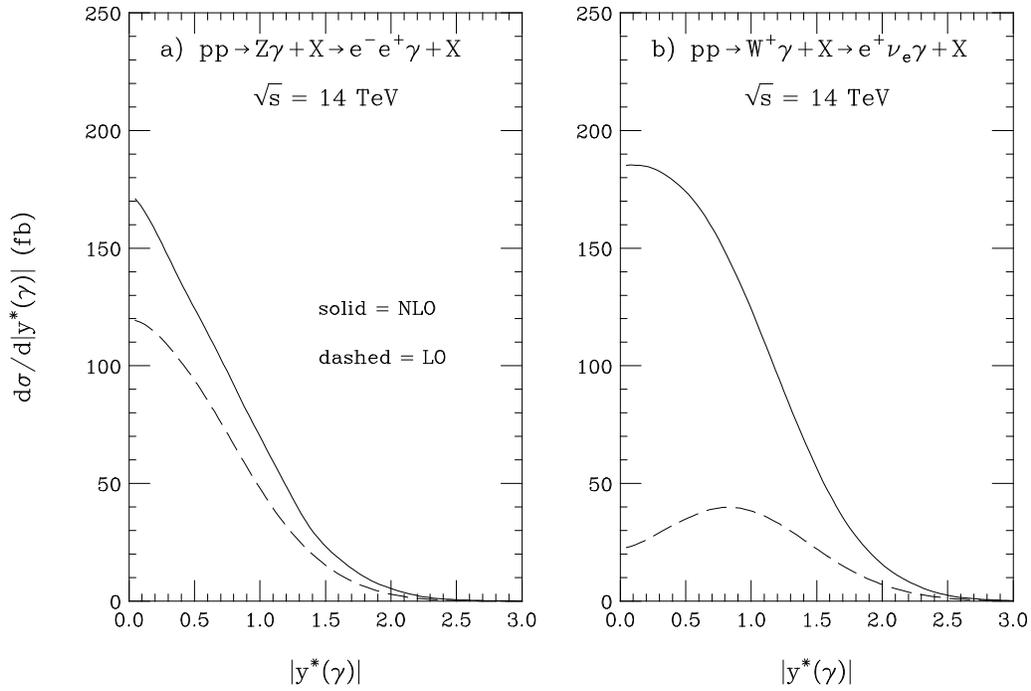,width=6.0in,clip= }
\caption{Same as Fig.~5, but for the LHC energy.}
\end{figure}

\begin{figure} 
\psfig{file=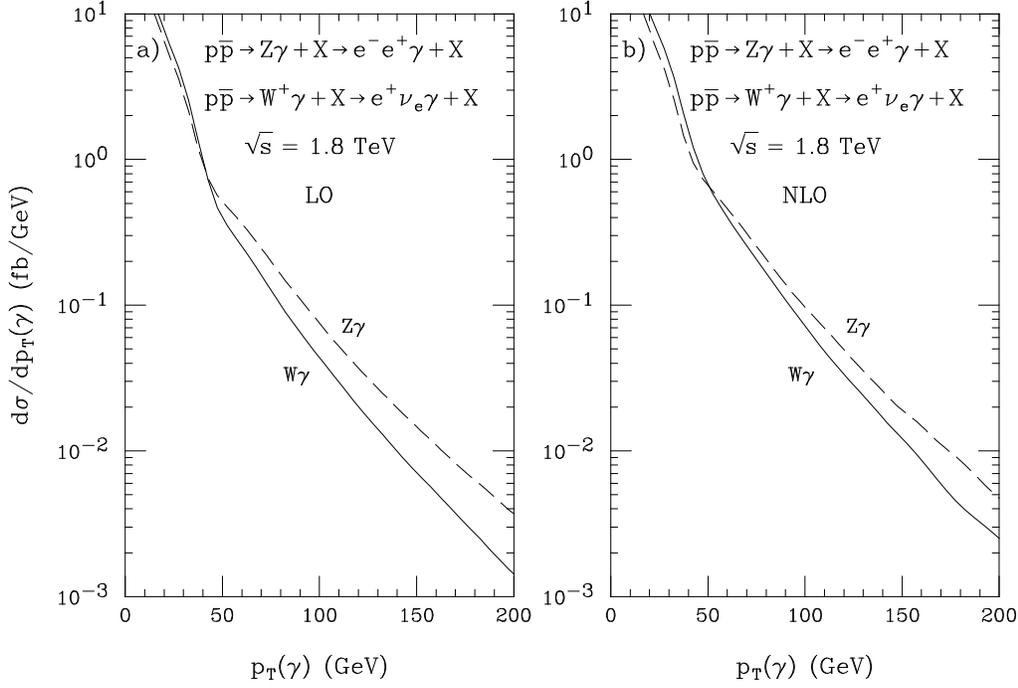,width=6.0in,clip= }
\caption{Photon transverse momentum distributions for $Z\gamma$ and $W\gamma$
production at the Tevatron energy.
Parts (a) and (b) are the LO and NLO cross sections, respectively.}
\end{figure}

\begin{figure} 
\psfig{file=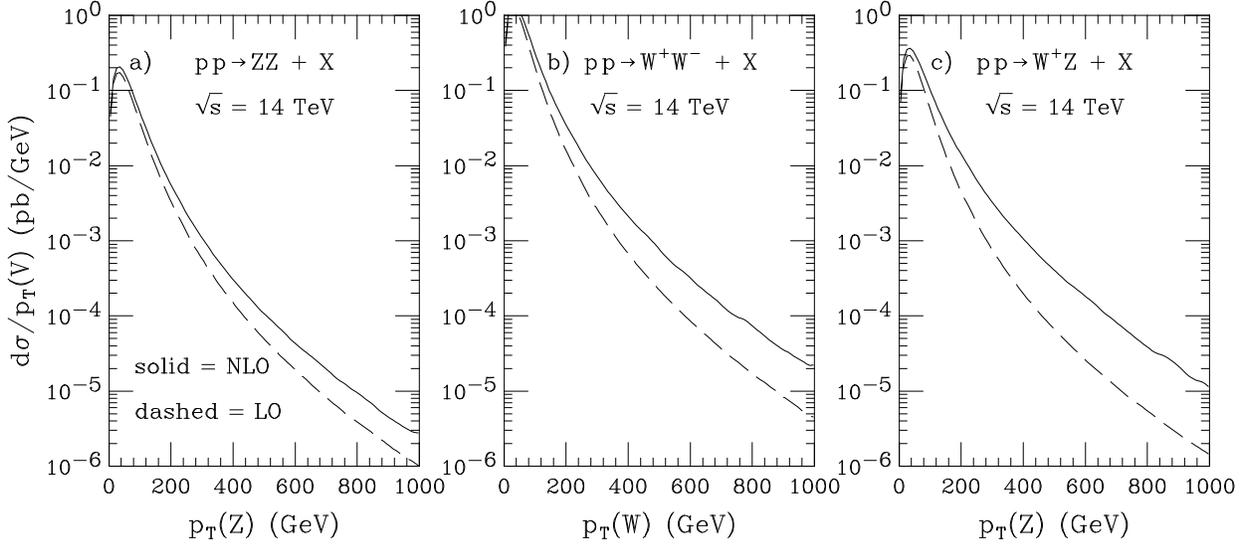,width=7.0in,clip= }
\caption{Weak boson transverse momentum distributions for
(a) $ZZ$, (b) $W^+W^-$, and (c) $W^+ Z$ production at the LHC energy.}
\end{figure}

\begin{figure} 
\psfig{file=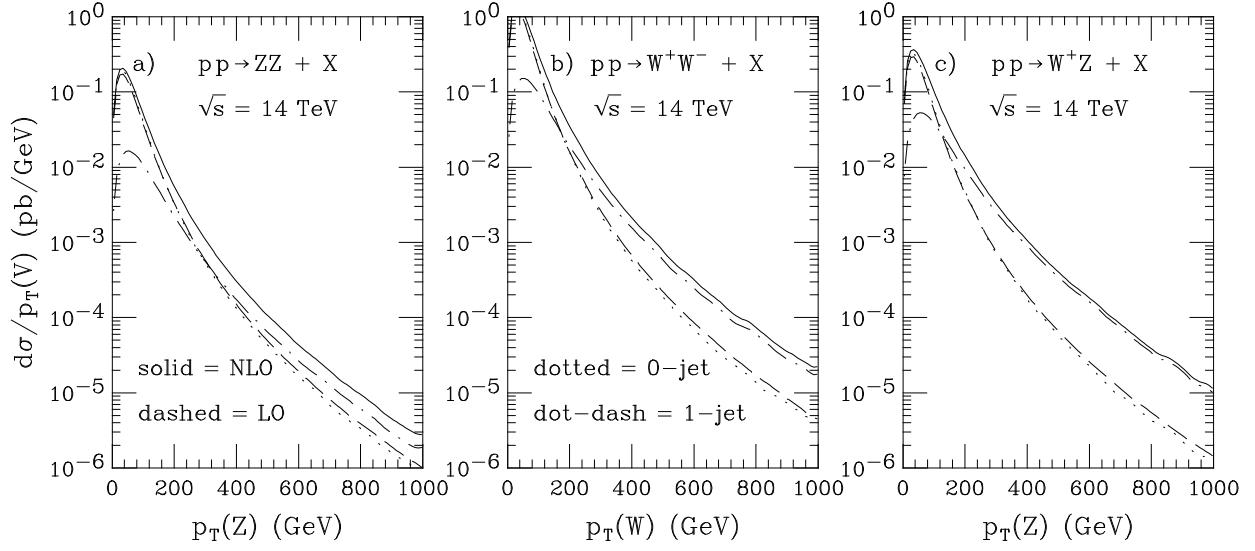,width=7.0in,clip= }
\caption{Same as Fig.~8. but now the 0-jet and 1-jet exclusive
components of the NLO inclusive cross section are also shown.}
\end{figure}

\begin{figure} 
\centerline{\psfig{file=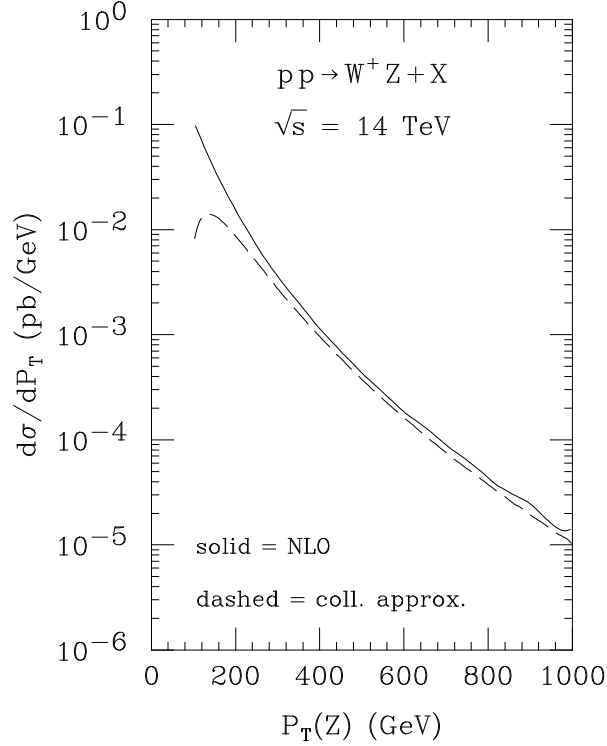,height=4.00in,clip= }}
\caption{The $p_T(Z)$ distribution for $pp \to W^+ Z + X$
at the LHC energy.
The full NLO cross section is compared to the
cross section obtained from the collinear
approximation given in Eq.~(1).}
\end{figure}

\begin{figure} 
\psfig{file=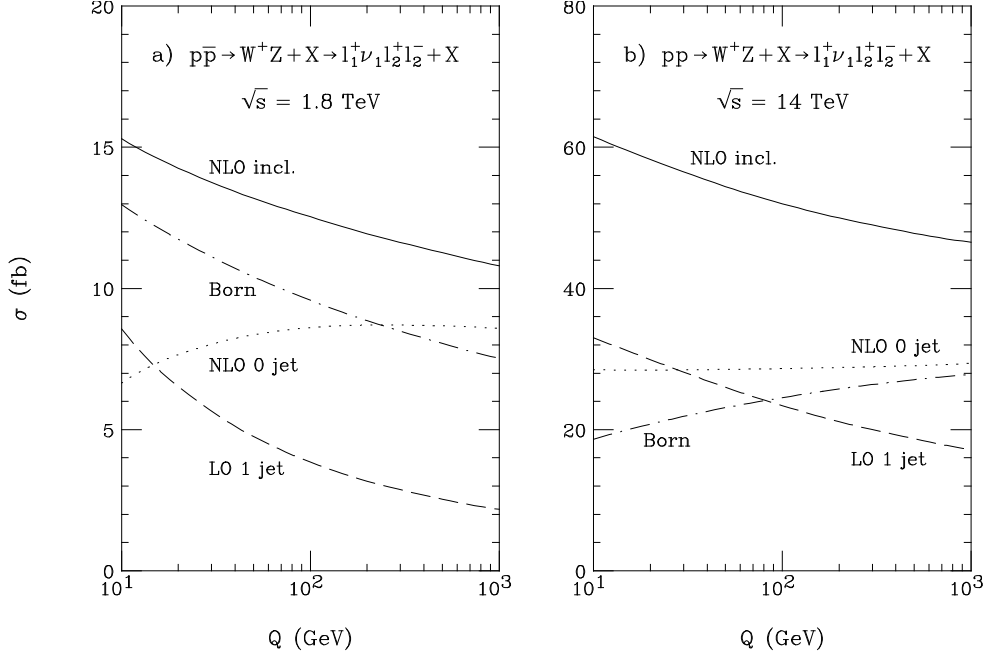,width=6.0in,clip= }
\caption{Total cross section for $W^+Z$ production as a function of the
scale $Q$ for (a) the Tevatron energy and (b) the LHC energy.
The Born, NLO inclusive, 0-jet exclusive, and 1-jet exclusive cross
sections are shown.}
\end{figure}

\begin{figure} 
\psfig{file=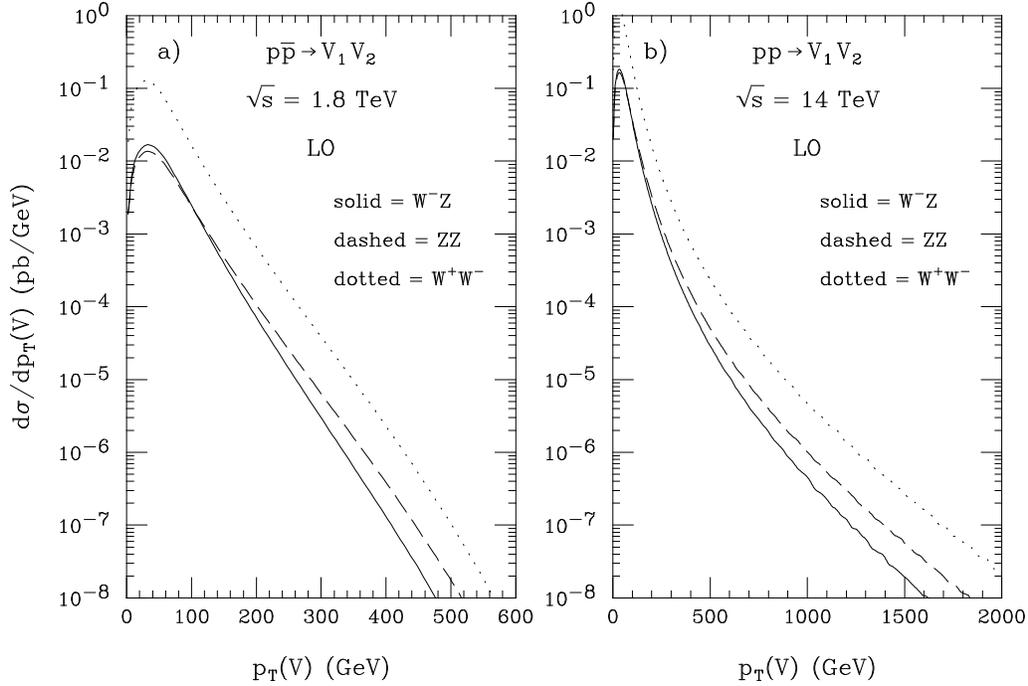,width=6.0in,clip= }
\caption{The weak boson transverse momentum distributions at LO for
$ZZ$, $W^+ W^-$, and $W^- Z$ production.
Parts (a) and (b) are for the Tevatron and LHC energies, respectively.}
\end{figure}


\begin{thebibliography}{99}

\bibitem{FIRSTWZ}
R.W.~Brown and K.O.~Mikaelian,
Phys. Rev. D {\bf 19}, 922 (1979);        
R.W.~Brown, K.O.~Mikaelian, and D.~Sahdev,
{\it ibid.} {\bf 20}, 1164 (1979).        
%
\bibitem{UCLA}
See talks by H.~Aihara, T.~Feuss, C.~Wendt, H.~Johari, L.~Zhang,
G.~Landsberg, B.~Wagner, and D.~Neuberger in these proceedings.  
%
\bibitem{EWSB}
D.~Dicus and V.~Mathur, Phys. Rev. D {\bf 7}, 3111 (1973);
M.~Veltman, Acta Phys. Pol. {\bf B8}, 475 (1977);
B.W.~Lee, C.~Quigg, and H.~Thacker, Phys. Rev. D {\bf 16}, 1519 (1977);
J.~van der Bij and M.~Veltman, Nucl. Phys. {\bf B231}, 205 (1984);
M.~S.~Chanowitz and M.~K.~Gaillard,
Nucl. Phys. {\bf B216}, 379 (1985).    
%
\bibitem{NLOVV}
J.~Ohnemus and J.F.~Owens,
Phys. Rev. D {\bf 43}, 3626 (1991);	
J.~Ohnemus, {\it ibid.}
{\bf 44}, 1403 (1991);	
{\bf 44}, 3477 (1991);	
{\bf 47}, 940 (1993). 	
%
\bibitem{NEWJO}
J.~Ohnemus,
Phys. Rev. D {\bf 50}, 1931 (1994);  
{\it ibid.}  {\bf 51}, 1068 (1995).  
%
\bibitem{BHO}
U.~Baur, T.~Han, and J.~Ohnemus,
Phys. Rev. D {\bf 48}, 5140 (1993).  
%
\bibitem{RAZ}
K.O.~Mikaelian, M.A.~Samuel, and D.~Sahdev,
Phys. Rev. Lett. {\bf 43}, 746 (1979);
R.W.~Brown, K.O.~Mikaelian, and D.~Sahdev
Phys. Rev. D {\bf 20}, 1164 (1979);
D.~Zhu, Phys. Rev. D {\bf 22}, 2266 (1980);
T.R.~Grose and K.O.~Mikaelian,
Phys. Rev. D {\bf 23}, 123 (1981);
C.J.~Goebel,  F.~Halzen, and J.P.~Leveille,
Phys. Rev. D {\bf 23}, 2682 (1981);
S.J.~Brodsky and R.W.~Brown, Phys. Rev. Lett. {\bf 49}, 966 (1982);
M.A.~Samuel, Phys. Rev. D {\bf 27}, 2724 (1983);
R.W.~Brown, K.L.~Kowalski, and S.J.~Brodsky,
Phys. Rev. D {\bf 28}, 624 (1983);
R.W.~Brown and K.L.~Kowalski,
Phys.~Rev. D {\bf 29}, 2100 (1984). 
%
\bibitem{FRIXWZ}
S.~Frixione, P.~Nason, and G.~Ridolfi,
Nucl. Phys. {\bf B~383}, 3 (1992).  
%
\bibitem{UJH}
U.~Baur, T.~Han, and J.~Ohnemus,
Phys. Rev. Lett. {\bf 72}, 3941 (1994).   
%
\bibitem{HAN}
T.~Han, in these proceedings.		  
%
\end{thebibliography}
\end{document}